\documentclass[twocolumn, numberedappendix]{aastex62}

\usepackage{graphicx}
\usepackage{epstopdf}
\usepackage{epsfig}
\usepackage{subfigure}
\usepackage{hyperref}
\usepackage{bm}
\usepackage{natbib}
\usepackage{color}
\usepackage[mediumspace,Gray,squaren]{SIunits}
\usepackage{multirow}
\usepackage{comment}
\usepackage{amsmath}
\usepackage[pass]{geometry}
\usepackage{aas_macros}
\usepackage[T1]{fontenc}
\usepackage{aecompl}
\usepackage{comment}

\includecomment{todo}
\begin{document}

\shorttitle{Accounting for selection bias using simulations}
\title{Accounting for selection bias using simulations: A general method and an application to millimeter-wavelength surveys}

\author[0000-0001-9032-1585]{Megan~B.~Gralla}
\affiliation{Department of Astronomy/Steward Observatory, University of Arizona, 933 N. Cherry Ave., Tucson, AZ 85721, USA}
\author[0000-0003-4496-6520]{Tobias~A.~Marriage}
\affiliation{Dept. of Physics and Astronomy, Johns Hopkins University, 3400 N. Charles St., Baltimore, MD 21218, USA}

\date{Accepted version}

\def\LaTeX{L\kern-.36em\raise.3ex\hbox{a}\kern-.15em
    T\kern-.1667em\lower.7ex\hbox{E}\kern-.125emX}

\begin{abstract}
We have developed a new Bayesian method to correct the flux densities of astronomical sources. The hybrid method combines a simulated likelihood to model survey selection together with an analytic source-count-based prior. The simulated likelihood captures the effect of complicated selection methods, such as multi-frequency filtering or imposed restrictions on recovered sample properties (e.g., color cuts). Simulations are also able to capture unanticipated sources of uncertainty. In this way, the method enables a broader application of Bayesian techniques. Use of an analytic prior allows variation of assumed source count models without re-simulating the likelihood. We present the method along with a detailed description of an application to real survey data from the Atacama Cosmology Telescope. 
\end{abstract} 


\section{Introduction}
\label{introduction}

Selecting astronomical sources from survey data in the presence of noise introduces biases in the brightness estimation of these sources and associated population statistics, a problem long-recognized in astronomy \citep[e.g.,][]{eddington1913,jauncey1968,murdoch1973,schmitt86,hogg1998,coppin2005}. Correcting flux density bias is especially critical in millimeter-wave surveys. The differential source counts of millimeter-wave selected source samples fall steeply with increasing flux density.  This tends to bias the measured flux density of a detected source, such that a given measured flux density is likely to correspond to a fainter intrinsic flux density that is boosted by a positive noise fluctuation.  Correcting the measured flux densities for this bias is referred to as ``deboosting'' the flux densities. 

Many recent studies of millimeter and infrared sources \citep[i.e., ][]{vieira2010, oliver2010, mocanu2013, marsden, valiante2016} have used the Bayesian formalism for deboosting developed in \citet{crawford2010} (hereafter C10).\footnote{An early example of the Bayesian framework applied to source flux bias is given in \cite{jauncey1968}.} For surveys with more than one source per resolution element, the interpretation of a measured flux density is ambiguous.  The C10 formalism has the advantage of providing a posterior distribution for the flux density of a single source: the brightest source in the resolution element. It accomplishes this by formulating a prior distribution for the intrinsic flux density of the brightest source that can be computed without simulations tailored to a particular survey. C10 also laid out the formalism for extending this approach to multiple bands with correlated prior distributions, with \citet{mocanu2013} extending the methods and applying them to three-band data, after \citet{vieira2010} had performed the deboosting of two-band data. 

The posterior distribution of the flux density can alternatively be calculated through a Monte Carlo simulation. For instance, \cite{coppin2005,coppin06} used simulations to compute a prior probability distribution for the underlying filtered flux density field given a source population. This simulated prior is then used to convert the analytic likelihood of the underlying flux density (given by the measured flux density and its error) into a posterior probability. This approach was adopted by \cite{austerman2009,austerman10} and \cite{marriagesources}. The principal difference between \cite{coppin2005,coppin06} and C10 is the formulation of this prior (and thus posterior). An analogous, but fully numerical, approach  involves generating maps and adding sources drawn from the prior counts distribution and spectral energy distributions. The simulated flux measurements can then be inverted to infer the underlying intrinsic flux distribution associated with a real, measured flux. These techniques have the ability to deal with unusual noise or selection properties. They are, however, computationally expensive. \cite{valiante2016} used simulations to compute the conditional probability of measured flux density given an intrinsic flux density, which was then used  to correct Eddington bias in source counts. However they followed the more tractable C10 method for flux deboosting. 
 
As noted above, the Monte Carlo methods for flux deboosting are computationally expensive, making them impractical to re-calculate for a large number of prior assumptions. In the posterior calculation, the likelihood contains the complications in the data and source selection, so benefits from a Monte Carlo approach. In contrast, the prior connects to a model and can usually be represented analytically. We take a hybrid approach to finding the posterior flux density distribution that combines a Monte Carlo likelihood estimation (to account for complexity in the map) with an analytic prior (to quickly test sensitivity to the prior). 

When a prior such as that developed by C10 is adopted, the ability to describe the intrinsic source flux density of the brightest source in a pixel is preserved. In addition, our Monte Carlo likelihood 1) uses the noise in the real maps rather than a model of the noise in the maps, 2) uses the full photometry pipeline of the actual survey (which could do something complex) rather than an analytic approximation to it. Some specific examples where the Monte Carlo can improve on analytic approximations include capturing all noise sources present in the data and correcting for selection effects in the likelihoods that may be difficult to model. In fact, we developed these techniques to account for selection effects of sample cuts due to Galactic contamination and multi-frequency filtering in maps from the Atacama Cosmology Telescope \citep[ACT; see ][]{swetz2011}.\footnote{Our method assumes that, once contamination (e.g., from cirrus) is culled from the data, the  noise (e.g., from detectors or background sources) is reasonably stationary across the map. If the map noise is significantly different in different regions, those regions will need to be treated (simulated, etc) separately.} In this paper, we will refer to our technique as ``debiasing'' because in this more general, multi-frequency formulation, flux densities may be corrected either up or down. There is a companion paper to this work \citep{grallasources} that presents ACT catalogs debiased using the method presented here, and \citet{datta} also use methods adapted from these.

This paper is organized as follows. In Section \ref{section:methods}, we describe the debiasing method to correct measured flux densities. In Section \ref{sec:implementation}, we demonstrate this method and discuss implementation details with examples using data from ACT. In Section \ref{sec:conclusion} we conclude. 

\section{Methods}\label{section:methods}
We build our methods off the Bayesian formalism described by C10. The posterior probability distribution $P(S_{i}|S_{m})$ for the intrinsic flux density of the brightest source within a resolution element ($S_{i}$) is
\begin{equation}\label{eqn:bayesdebias}
P(S_{i}|S_{m}) \propto P(S_{m}|S_{i})P(S_{i}), 
\end{equation}
where $S_{m}$ is the total flux density measured in that resolution element\footnote{The relevant resolution element is related to the resolution of the observations rather than any scale related to map pixelization choices.} (or ``pixel''); $P(S_{m}|S_{i})$ is the conditional probability of measuring flux $S_{m}$ in a pixel given that the brightest source within it has intrinsic flux density $S_{i}$, or, viewed as a function of $S_i$, the likelihood of $S_i$ given the measured data; and $P(S_{i})$ is the prior probability that the brightest source in that pixel has flux density $S_{i}$.  The prior probability distribution is determined by a model for the source counts (proportional to the probability that a source of flux $S_i$ is in the pixel) modified by the exponential Poissonian probability that there are zero sources brighter than $S_i$ in the pixel, as in C10:
\begin{equation}\label{eqn:prior}
P(S_{i}) \propto \frac{dN}{dS} \exp\bigg(-\Delta \Omega_{p} \int_{S_{i}}^{\infty} \frac{dN}{dS'} dS' \bigg),
\end{equation}
where $\Delta \Omega_{p}$ is the pixel area and $\frac{dN}{dS}$ is the differential number counts, expressed per unit flux density per unit solid angle. In practice the pixel area can be taken as the square of the full width half maximum of a telescope beam \citep{crawford2010}.

\begin{figure*}
	\centering
	\includegraphics[width=110mm]{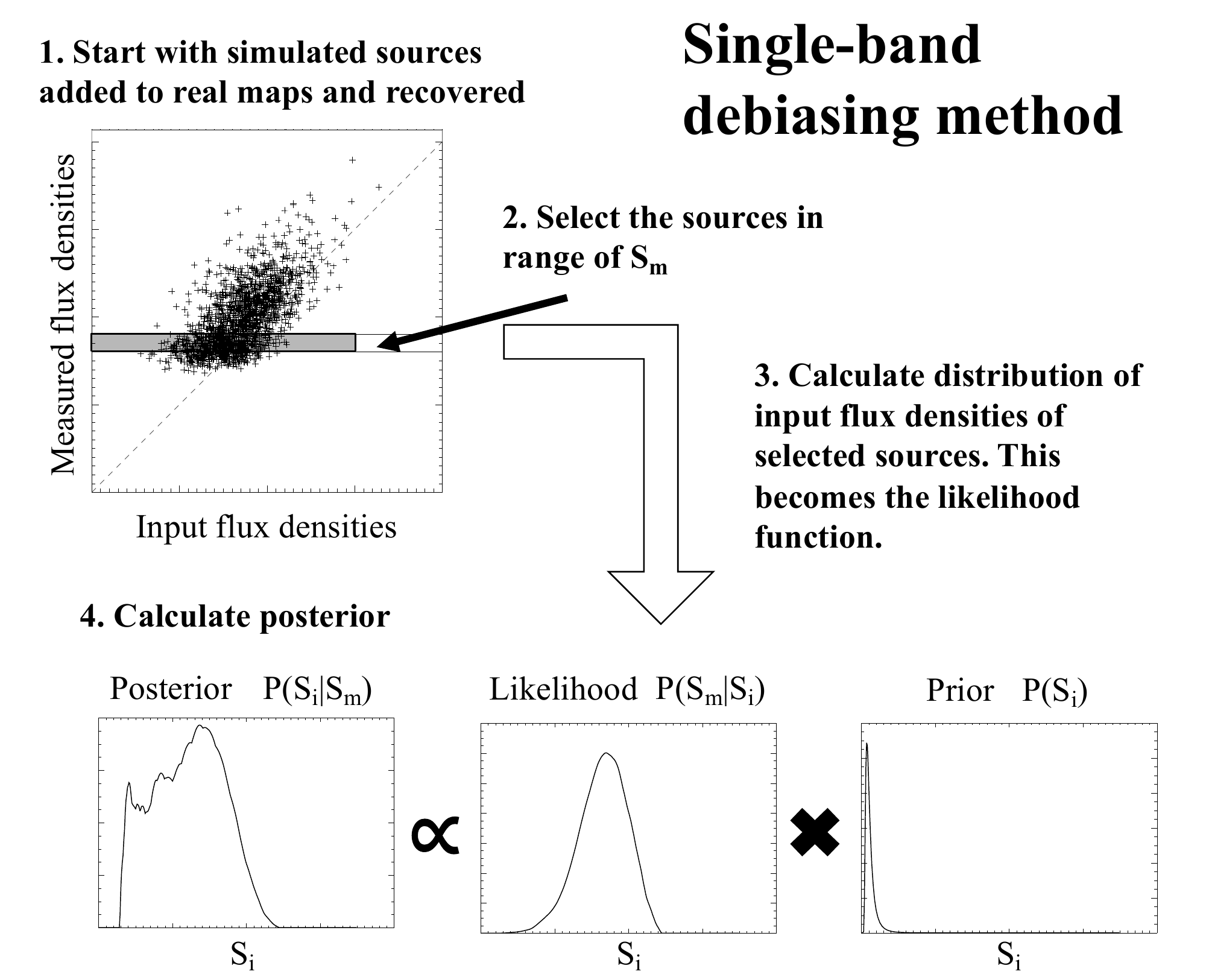}
\caption{Summary of methods for single-band debiasing. We begin by adding simulated sources to the maps, which include noise as well as signal from the CMB, sources, and the Galaxy. We apply the same selection algorithm as used to generate the real catalogs. We select the recovered simulated sources with measured flux density  similar to the flux density ($S_{m}$) of the source that we would like to debias. The distribution of input flux densities of these selected simulated sources becomes the likelihood function. Multiplying the likelihood function by an analytic prior (Equation \ref{eqn:prior}) allows us to calculate the posterior probability distribution of the intrinsic flux density of the brightest source within a pixel ($S_i$) corresponding to $S_m$. We calculate the debiased flux density from this posterior distribution.
\label{summary}}
\end{figure*}

In contrast to previous approaches, which rely on approximate analytic formulations, we numerically estimate the likelihood of $S_i$ by
adding simulated sources to the real maps, recovering the sources, and comparing the input and recovered flux densities. 
We then sample the simulation along slices of fixed $S_m$ by calculating the density distribution of the simulated sources as a function of $S_i$, at fixed $S_m$. The resulting function is the likelihood.\footnote{Note that the likelihood of the $S_i$ is derived from the conditional distribution of $S_m$ given  $S_i$. Therefore the likelihood   is not a distribution of $S_i$, but rather a function of $S_i$, which through Bayes' Theorem allows us to derive the  distribution of $S_i$.} Once simulated, the likelihood function can then be multiplied by any number of prior distributions, which depend on assumptions about the underlying source population. 

In practice, for a given measured flux density, we identify sources recovered from the simulations that have a similar measured flux density within some tolerance. We identify the input flux densities of these sources, and we estimate the number density of these recovered simulated sources as a function of the input flux density using a kernel density estimator. Note that the likelihood is typically understood as a distribution of  measured values, but here we calculate it as a function of intrinsic values, at fixed measured flux density. This lends itself more directly to our goal of the determination of the posterior as a function of the intrinsic flux density.
We then multiply this likelihood function  by the  prior (Equation \ref{eqn:prior}) to calculate the posterior probability distribution, properly normalized, of intrinsic flux density of the brightest source given a measured flux density (Equation \ref{eqn:bayesdebias}).  The median and 68\% confidence interval of this posterior probability distribution are reported as the debiased flux densities.  
Figure \ref{summary} provides a pictorial summary of these methods described for debiasing flux densities in a single band.

\subsection{Two-Band Debiasing}

In principle, the single-band approach is straightforward to extend to two bands. In this case the posterior is
\begin{align}
    P(S_{i,1},S_{i,2} & \mid S_{m,1},  S_{m,2}) \nonumber \\ & \propto P(S_{m,1},S_{m,2} \mid S_{i,1}, S_{i,2}) P(S_{i,1}, S_{i,2}),
    \label{eqn:twobandposterior}
\end{align}
where subscripts now differentiate between bands as well as intrinsic and measured specifications. The function ${P(S_{m,1},S_{m,2}\mid S_{i,1}, S_{i,2})}$ is a simulated two-band likelihood, and ${P(S_{i,1}, S_{i,2})}$ is an analytic prior. This prior can be rewritten as ${P(S_{i,2}\mid S_{i,1}) P(S_{i,1})}$. Here ${P(S_{i,1})}$ is the single-band prior (Equation \ref{eqn:prior}). The two flux densities are related by a spectral index $\alpha$ such that
\begin{equation}\label{eqn:alpha}
(S_{i,2}) = S_{i,1}(\nu_{2}/\nu_{1})^{\alpha}
\end{equation}
where $\nu_{2}$ and $\nu_{1}$ refer to the secondary and primary band frequencies, respectively.
There is some distribution for $\alpha$, which allows one to approximate the conditional probability ${P(S_{i,2}\mid S_{i,1})}$. 

An alternative to this formulation, given in C10, re-parameterizes the model in Equation \ref{eqn:twobandposterior} in terms of $S_{i,1}$ and $\alpha$. In this case the prior ${P(S_{i,1},\alpha)}$ can be rewritten as ${P(\alpha \mid S_{i,1}) P(S_{i,1})}$. Here again ${P(S_{i,1})}$ is given by Equation \ref{eqn:prior}. The conditional probability ${P(\alpha \mid S_{i,1})}$ is the normalized sum of spectral index distributions for different source types (e.g., AGN and DSFGs in the ACT data), weighted by the number of expected sources of each type, as established by count models for $S_1$. The resulting posterior ${P(S_{i,1},\alpha \mid S_{m,1},  S_{m,2})}$ can be marginalized to give the single parameter posterior probability for $S_{i,1}$, $\alpha$ and, after a transformation of variables, $S_{i,2}$.

\begin{figure*}
	\centering
	\includegraphics[width=\textwidth]{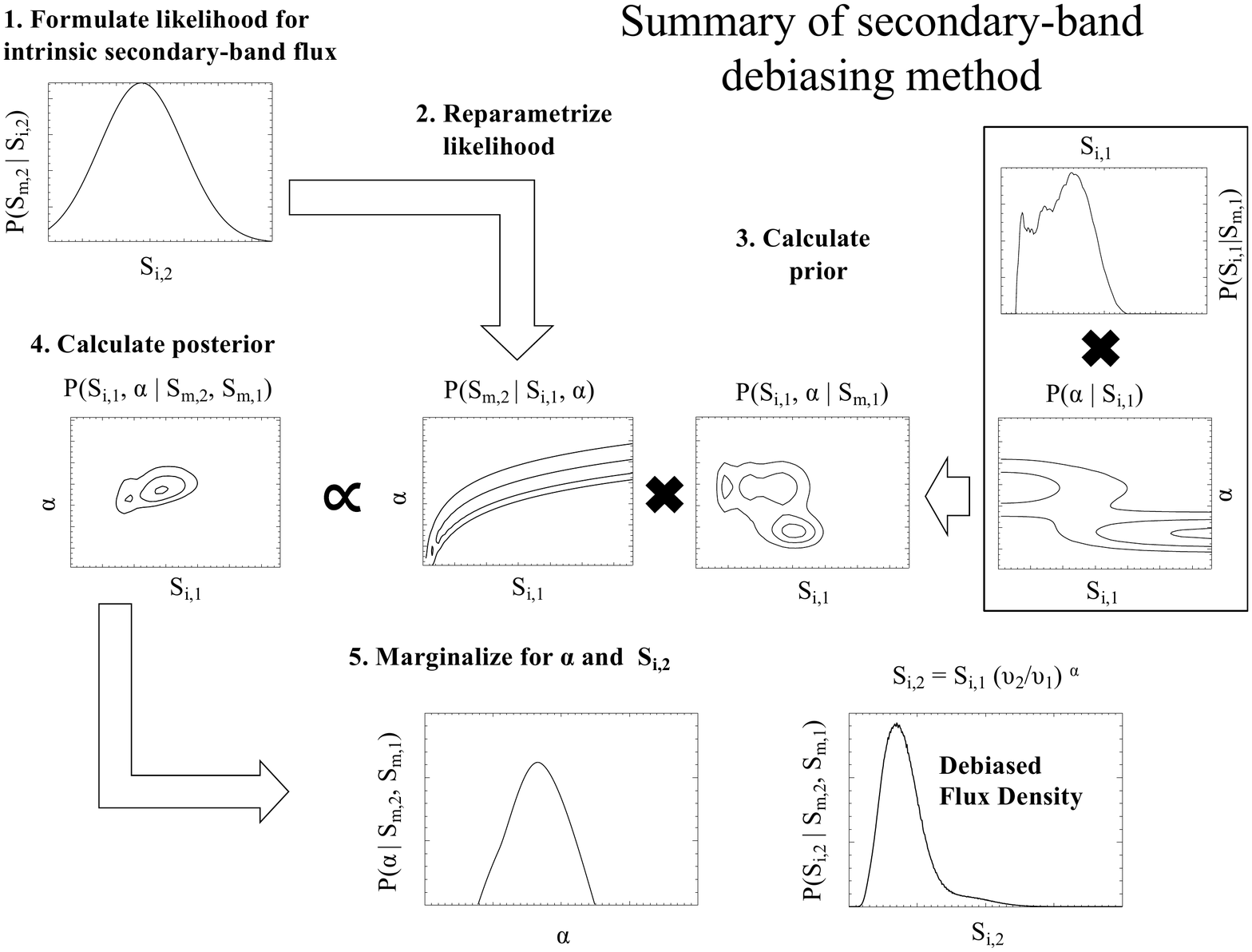}
\caption{Summary of methods for debiasing flux densities of multiple bands. As in the single-band case, we start with simulated sources that have been recovered in the primary band and also measured in the secondary band. We start with the likelihood for $S_{i,2}$ based on the Gaussian probability distribution for $S_{m,2}$. We then express this likelihood as a function of $S_{i,1}$ and $\alpha$. To construct the prior as a function of $S_{i,1}$ and $\alpha$, we provide the conditional distribution for spectral index given $S_{1,i}$. The prior on $S_{i,1}$ is given by the posterior distribution calculated according to the single-band debiasing summarized in Figure~\ref{summary}. We multiply the resulting prior probability density distribution by the likelihood function to determine the posterior probability density as a function of $S_{i,1}$ and $\alpha$. This is then tranformed and marginalized to give the posterior distribution of $S_{i,2}$ using Equation~\ref{eqn:bayesdebias2dto1d}. We calculate the debiased flux density in the secondary band from this posterior distribution.
\label{summary2}}
\end{figure*}

In practice, producing simulations comprehensive enough for computing the full two-band likelihood in Equation \ref{eqn:twobandposterior} can be a challenge. This is the case with the ACT data where the required flux densities spanned two orders of magnitude ($\sim1-100$~mJy) with a broad range of spectral indices. In \cite{grallasources} we implement a simplified version that captures the important aspects of debiasing while proving more tractable computationally. 

The approach can be described as follows:
\begin{align}
    P(&S_{i,1},\alpha  \mid  S_{m,1},  S_{m,2})  \propto P(S_{m,1}, S_{m,2} \mid S_{i,1},\alpha) P(S_{i,1},\alpha) \nonumber \\
    & \approx P(S_{m,1} \mid S_{i,1}) P(S_{m,2} \mid S_{i,1}, \alpha)P(\alpha \mid S_{i,1}) P(S_{i,1}) \nonumber \\
    & \propto P(S_{m,2} \mid S_{i,1}, \alpha)  P(\alpha \mid S_{i,1}) P(S_{i,1} \mid S_{m,1})   \nonumber\\
    & =  P(S_{m,2} \mid S_{i,1}, \alpha)  P(S_{i,1}, \alpha \mid S_{m,1}).   
    \label{eqn:bayesdebias2d}
\end{align}
In the second line we make the approximation that the likelihood function of the two bands can be decomposed into independent likelihoods, $P(S_{m,1} \mid S_{i,1})$ and $P(S_{m,2} \mid S_{i,1}, \alpha)$. In the third line we have combined the primary-band likelihood $P(S_{m,1} \mid S_{i,1})$ with the primary-band prior $P(S_{i,1})$ to produce the primary-band posterior distribution (Equation \ref{eqn:bayesdebias}). In the final line, we collapse the product of the $S_{i,1}$ posterior and conditional $\alpha$ distribution into a joint conditional prior $P(S_{i,1}, \alpha \mid S_{m,1})$ on $S_{i,1}$ and $\alpha$. The secondary-band likelihood $P(S_{m,2} \mid S_{i,1}, \alpha)$ can be taken as a Gaussian approximation given the raw flux density and error or can be simulated in its own right. Because the source selection is carried out in the primary band, simulating this band is most important to capture selection effects in the likelihood. Conversely, secondary band simulations are less useful unless the secondary band also plays a significant role in source selection.

In practice, we calculate this method numerically on one or two-dimensional grids.
For instance, to calculate the marginalized probability distribution  for $S_{i,2}$ starting from Equation~\ref{eqn:bayesdebias2d}, each $(j,k)$th pixel location in a two-dimensional grid of ($S_{i,1},\alpha$) is evaluated for its corresponding secondary band flux density according to Equation \ref{eqn:alpha}.
The one-dimensional probability density array for $S_{i,2}$ is computed as
\begin{equation}\label{eqn:bayesdebias2dto1d}
P(S_{i,2}\mid S_{m,1},S_{m,2})_\ell \propto \sum_{j,k}^{n} P(S_{i,1},\alpha \mid S_{m,2}, S_{m,1})_{j,k},
\end{equation}
where the sum is performed over the $n$ $(j,k)$ pixels that satisfy $(S_{i,2})_\ell < (S_{i,1})_j (\nu_{2}/\nu_{1})^{\alpha_k} < (S_{i,2})_{\ell+1}$. 
The debiased secondary band flux density can then be calculated from the median (and the uncertainties from the 68\% confidence intervals) of this posterior distribution evaluated for all $\ell$ bins, $P(S_{i,2}\mid S_{m,1},S_{m,2})_\ell$. Figure \ref{summary2} provides a pictorial summary of these debiasing methods for the flux density of secondary bands. 

These methods can be generalized to samples selected by combining multiple bands, as in the case of multi-frequency matched filters. They can also be applied to samples selected from different electromagnetic regimes altogether (such as through the relation between far-infrared and radio). The main requirements are that there is some prior information about the selection that is included (such as the source counts), and that the selection be simulated along with the determination of the flux densities in all bands of interest.

\section{Implementation} \label{sec:implementation}
In this Section, we demonstrate these methods for the debiasing of millimeter sources using simulated sources added to data from the Atacama Cosmology Telescope. 

\subsection{Constructing the simulated source catalogs}\label{sec:sims}

We simulate observations in the 218~GHz ACT band. 
Because we are interested in the intrinsic flux densities for a population of dusty galaxies, we modeled our simulated source population to be like that expected for dusty galaxies. For each of eighteen trials, we generate 1,000 sources at random positions throughout the maps. Flux densities are randomly assigned according to a Gaussian distribution with mean 10~mJy and standard deviation 5~mJy. This was chosen to provide adequate statistics near the sensitivity limit of the survey, where debiasing is expected to be more important. Because the number of simulated sources drops off at the high end, we supplemented these simulations with 9 trials of simulations populated uniformly in flux density in the range $0-100$~mJy. Because the input flux density distribution was not a uniform random distribution, we normalize the likelihood function that we calculate from these simulations by the distribution of input flux densities. 

The 218~GHz flux densities are then scaled by randomly distributed spectral indices to assign flux densities to these sources in the 148 band. The input spectral index distribution is normally distributed as  $\alpha=3.4\pm1.3$, as based on the sample mean and standard deviation measured from the raw flux densities of the dusty galaxies of a preliminary version of the ACT equatorial catalog \citep{grallasources}. 

\begin{figure}
	\centering
	\includegraphics[width=84mm]{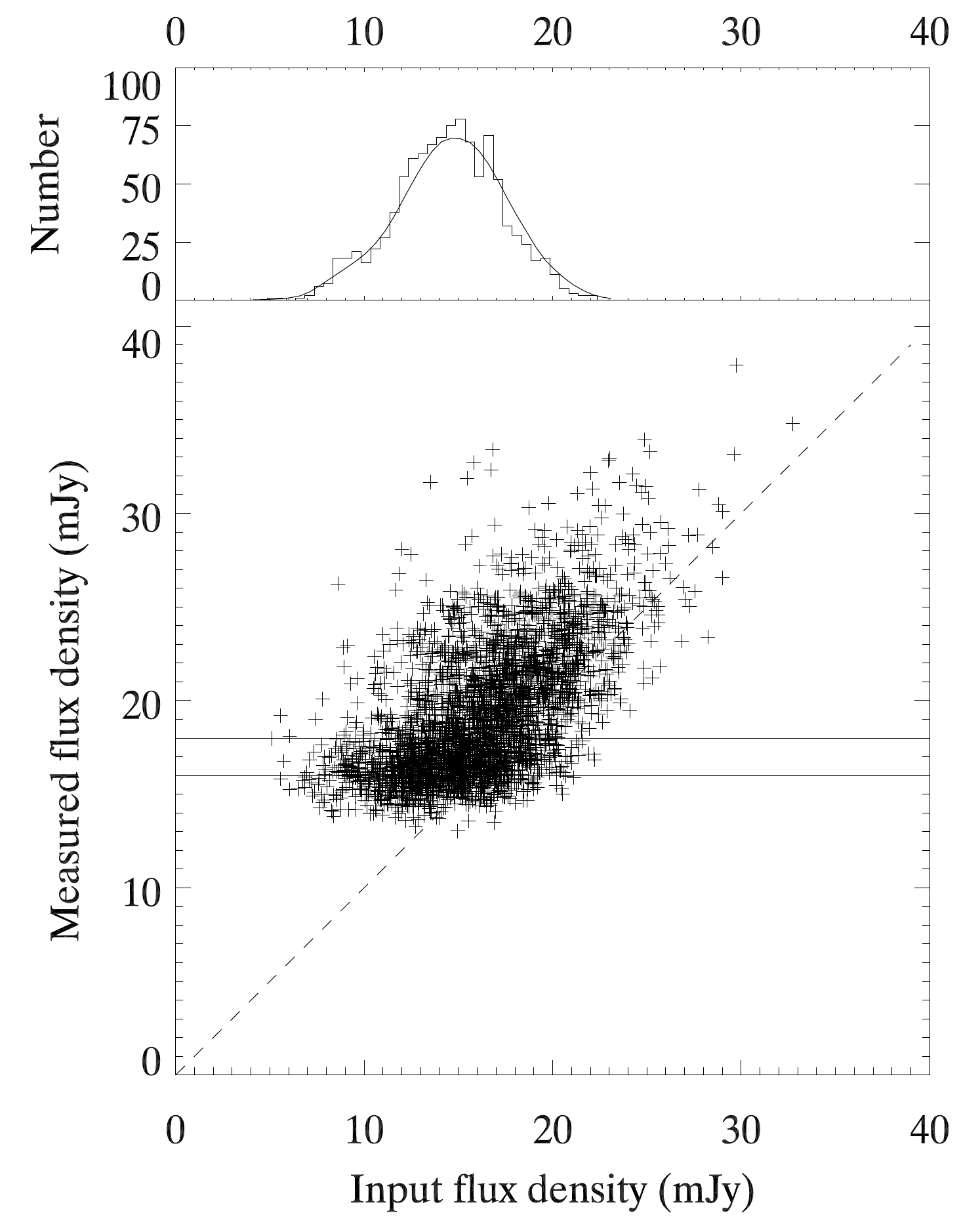}
\caption{Bottom: The input and measured flux densities of simulated sources that have been added to real 218~GHz data and recovered using our source selection methods. The dashed line indicates where the measured flux density equals the input flux density. (At this point in the simulation, we have not corrected the likelihood for the distribution of input flux densities. This accounts for why the measured flux densities as plotted here  lie below the input flux densities, as discussed in Section \ref{sec:sims}.) In order to populate the likelihood function, we sample these simulations along slices of constant measured flux density and estimate the resulting density distribution as a function of input flux density. The solid lines indicate the region from which we select simulated sources to calculate the likelihood function for debiasing the example source at $S_{218} = 17$~mJy, described in Section \ref{sec:example}. Top: Histogram of the simulated sources selected from the slice used to debias the example source. Also shown is the smoothed distribution from the kernel density estimator used to construct the likelihood function.}
\label{examplesinglebandsims}
\end{figure}

The simulated source template is then added to the map data, which include instrumental and atmospheric noise as well as signal from the cosmic microwave background, Galactic dust, and other compact sources such as AGN and dusty galaxies. Because we use a matched filter in our source recovery \citep[see for example, ][]{marriagesources}, we apply the same filter to both the source template and the data before combining them.
We then run the source selection method on the simulated data set and measure the recovered flux densities. Figure \ref{examplesinglebandsims} shows the input and recovered flux densities for the simulated source sample. Note that the distribution of measured flux densities tends to peak  below the input flux densities. This occurs because the input flux density distribution peaks at low (10~mJy) flux densities, so many more faint sources are input to the simulations. As discussed above, we remove this effect by normalizing the calculated likelihood function with the inverse of the input flux density distribution. We verified that the resulting median of the likelihood function agrees with the measured flux density across the full range of flux densities observed. 

\subsection{An example source} \label{sec:example}

For illustration, we walk through the debiasing of an example source drawn from the simulated sample of dusty galaxies described in Section \ref{sec:sims}. The measured $S_{m,148}$ and  $S_{m,218}$ flux densities are 4 and 17~mJy, respectively.   
Typically, the primary band is the band for which the S/N is highest. For this example, the primary band is 218~GHz, where the S/N would be about 6. 

\begin{figure}[ht]
	\centering
	\includegraphics[width=84mm]{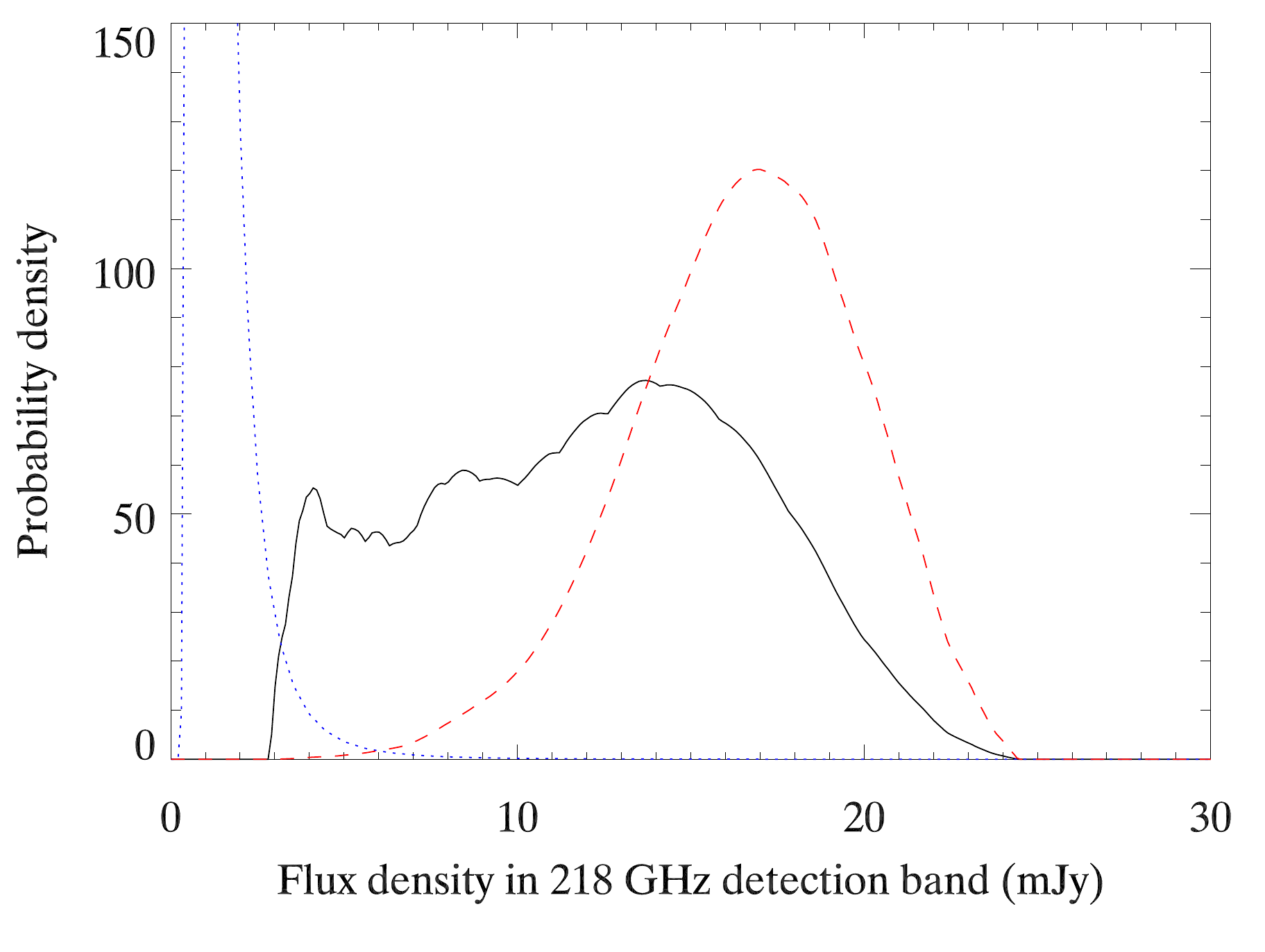}
\caption{The posterior $P(S_{i,218}\mid S_{m,218})$ (solid black), the likelihood $P(S_{m,218}\mid S_{i,218})$ (dashed red), and the prior $P(S_{i,218})$ (dotted blue) of the intrinsic 218~GHz flux density of a source with measured $S_{m,218}=17$~mJy.   \label{examplesingleband}}
\end{figure}

\subsubsection{Single or primary-band debiasing}\label{singleband}

To debias the flux density of the primary band (218~GHz for this example), we must calculate the likelihood function $P(S_{m}|S_{i})$.  We select sources from our simulations with a measured value of $S_{m,218}$ similar to the value that we are debiasing.  The default tolerance for 218~GHz selected sources is $\pm$1.0~mJy, so all simulated sources with measured $S_{m,218}$ between 16 and 18~mJy are included for this example. See Figure \ref{examplesinglebandsims} for illustration. We apply a kernel density estimator to the distribution of $S_{i,218}$ of these simulated sources. The resulting function of $S_{i,218}$ is the likelihood, shown in Figure~\ref{examplesingleband}.\footnote{For this figure, we used a larger tolerance ($\pm$1.5~mJy) to increase the statistics in order to better represent the underlying of the distribution. The quoted medians and uncertainty ranges are not affected by this choice.}  

We calculate the prior probability for $S_{i,218}$ according to Equation \ref{eqn:prior}. 
For the 218~GHz source counts, we construct a model from the addition of a model for AGN source counts from \citet{tucci} (the C2Ex model) and a model for dusty galaxies from \citet{bethermin2011}. The size of the resolution element $\Delta \Omega_p$, or the square of the full width at half maximum of the beam, for ACT is $1.6\times10^{-7}$~steradians at 148~GHz and $9.0\times10^{-8}$~steradians at 218~GHz \citep{hasselfieldbeam}.

Finally, the posterior probability distribution for $S_{i,218}$ is computed from the likelihood function and the prior (Equation \ref{eqn:bayesdebias}).  The posterior, the likelihood, and the prior for this example source are shown in Figure \ref{examplesingleband}.  The debiased flux density that we report at 218~GHz is the median of the posterior distribution, which for this example is 13.2~mJy.  The uncertainty is taken to be the $68\%$ confidence interval, which for this example source is $8.1-17.3$~mJy.

\subsubsection{Secondary band debiasing}

\begin{figure}
\centering
	\includegraphics[width=84mm]{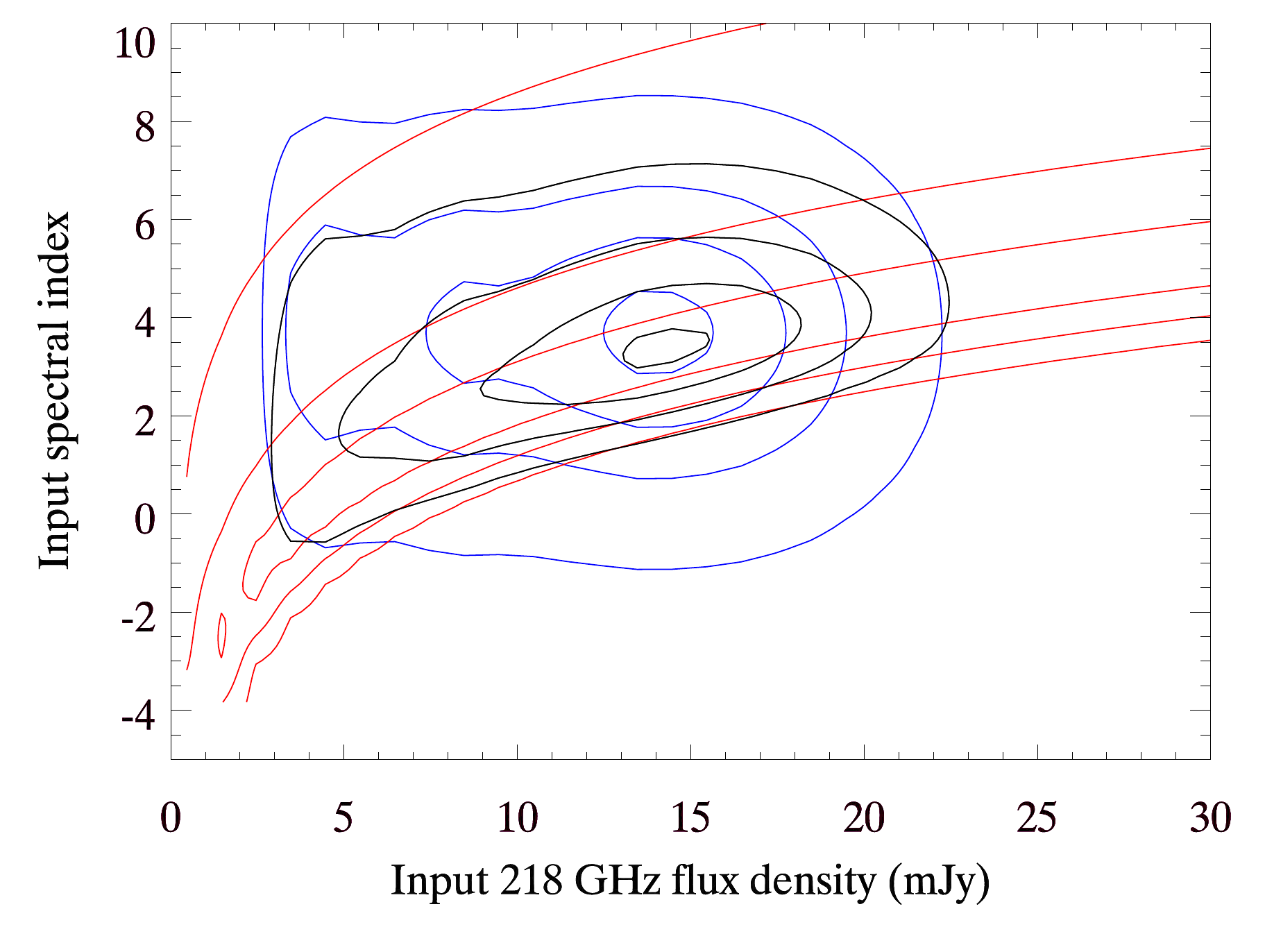}
\caption{The secondary-band likelihood  ${P(S_{m,148}\mid S_{i,218},\alpha)}$ (red),  prior  ${P(S_{i,218},\alpha \mid S_{m,218})}$ (blue), and  posterior   ${P(S_{i,218},\alpha \mid S_{m,148}}, S_{m,218})$ (black) for an example dusty galaxy with $S_{m,148}=4$~mJy and $S_{m,218}=17$~mJy (Equation~\ref{eqn:bayesdebias2d}).
\label{examplecontours}
}
\end{figure}

After thus debiasing the primary band flux density (218~GHz in this example), we then proceed to debias the secondary bands.   
We model the $S_{i,148}$ likelihood as a Gaussian with mean equal to the measured value and standard deviation equal to the uncertainty on the measurement. We then map this likelihood function to the $S_{i,218}-\alpha$ plane. In practice, for each pixel in a $S_{i,218}-\alpha$ grid, we assign the value ${P(S_{m,148} \mid S_{i,148})}$ where  $S_{i, 148}$ is re-expressed according to Equation \ref{eqn:alpha} as $S_{i, 218}(148/218)^{\alpha}$.
Figure \ref{examplecontours} shows the likelihood ${P(S_{m,148} \mid S_{i,218},\alpha)}$ for this example.  

The prior probability distribution ${P(S_{i,218},\alpha \mid S_{m,218})}$ factors into two parts. The first part is the posterior $P(S_{i,218} \mid S_{m,218})$ shown in Figure \ref{examplesingleband}. The second part of the prior is the conditional probability ${P(\alpha \mid S_{i,218})}$, which is computed as follows.  

For every pixel in the grid of primary band flux densities, we calculate the prior probability density distribution for $\alpha$.
For the prior probability of $\alpha$, we use source counts models \citep[][for AGN and DSFGs, respectively]{tucci,bethermin2011} to calculate the expected ratio of the number of AGN to the number of SFGs as a function of flux density. We model the probability density distribution of the spectral index for each population as Gaussians, for example:
\begin{equation}\label{eqn:alphagauss}
P_X(\alpha) = \frac{1}{\sigma_{X} \sqrt{2 \pi}} \exp\left(\frac{-(\alpha - \alpha_{X})^2}{2 \sigma_{X}^2}\right),
\end{equation}
where for subpopulation $X$ ($X=a$ for AGN and $X=d$ for DSFGs) $\alpha_{X}$ and $\sigma_{X}$ are the mean and standard deviation in $\alpha$ for that subpopulation, respectively.
The mean and standard deviation of each Gaussian is determined by calculating the sample median and standard deviation of the $\alpha$ distributions for the sources in our catalog (from the region of the map with less noise, and using their measured flux densities prior to debiasing), with the dividing line between AGN and DSFGs defined as $\alpha=1.0$. In order to make the prior on $\alpha$ somewhat less constraining of the resulting flux densities, we choose the standard deviation to be greater than that of the measured sample standard deviation. This widens the prior on $\alpha$, which when normalized then has less constraining power on the measurement-based likelihood function. For AGN, the prior we use has median $\alpha_a=-0.7$ and the standard deviation is $\sigma_a=1.2$ (the sample standard deviation is 0.9).  For the DSFGs, the median $\alpha_d=3.7$ and the standard deviation is $\sigma_d=2.2$ (the sample standard deviation is 1.6).\footnote{We tested our sensitivity to the width of the prior on $\alpha$ by also calculating the debiasing using the narrower sample standard deviations instead and comparing the results. We found that the majority of the sample is not affected, but that in particular, outliers classified as DSFGs that have weaker 148~GHz detections are more debiased when the prior $\alpha$ distribution is narrower. This behavior is expected: intuitively, the unchanged prior on the mean $\alpha$ implies that most of the sample should be unchanged, but for sources that deviate from the mean behavior, a tighter prior on $\alpha$ brings more statistical weight from the primary band to bear on the debiasing calculated for the secondary band. } 
We scale each of these distributions by the fraction of sources expected based on the counts models, and add the distributions together to get a distribution for $\alpha$ as a function of flux density:
\begin{equation}\label{eqn:alphaprior}
P(\alpha \mid S_{i,218}) \propto f_a(S_{i,218}) P_a(\alpha) + f_d(S_{i,218}) P_d(\alpha),
\end{equation}
where $f_a$ and $f_d$ are the fractions of AGN and DSFGs, respectively, as determined by the source counts models, and $P_a(\alpha)$ and $P_d(\alpha)$ are calculated according to Equation \ref{eqn:alphagauss}. 
For each $S_{i,218}-\alpha$ pixel, we multiply the conditional probability for $\alpha$ by the posterior probability for $S_{i,218}$:
\begin{equation}
P(S_{i,218},\alpha \mid P_{m,218}) \propto P(\alpha \mid S_{i,218}) P(S_{i,218} \mid P_{m,218}).
\end{equation}
The resulting prior is shown in blue in Figure \ref{examplecontours}.

\begin{figure}
	\centering
	\includegraphics[width=84mm]{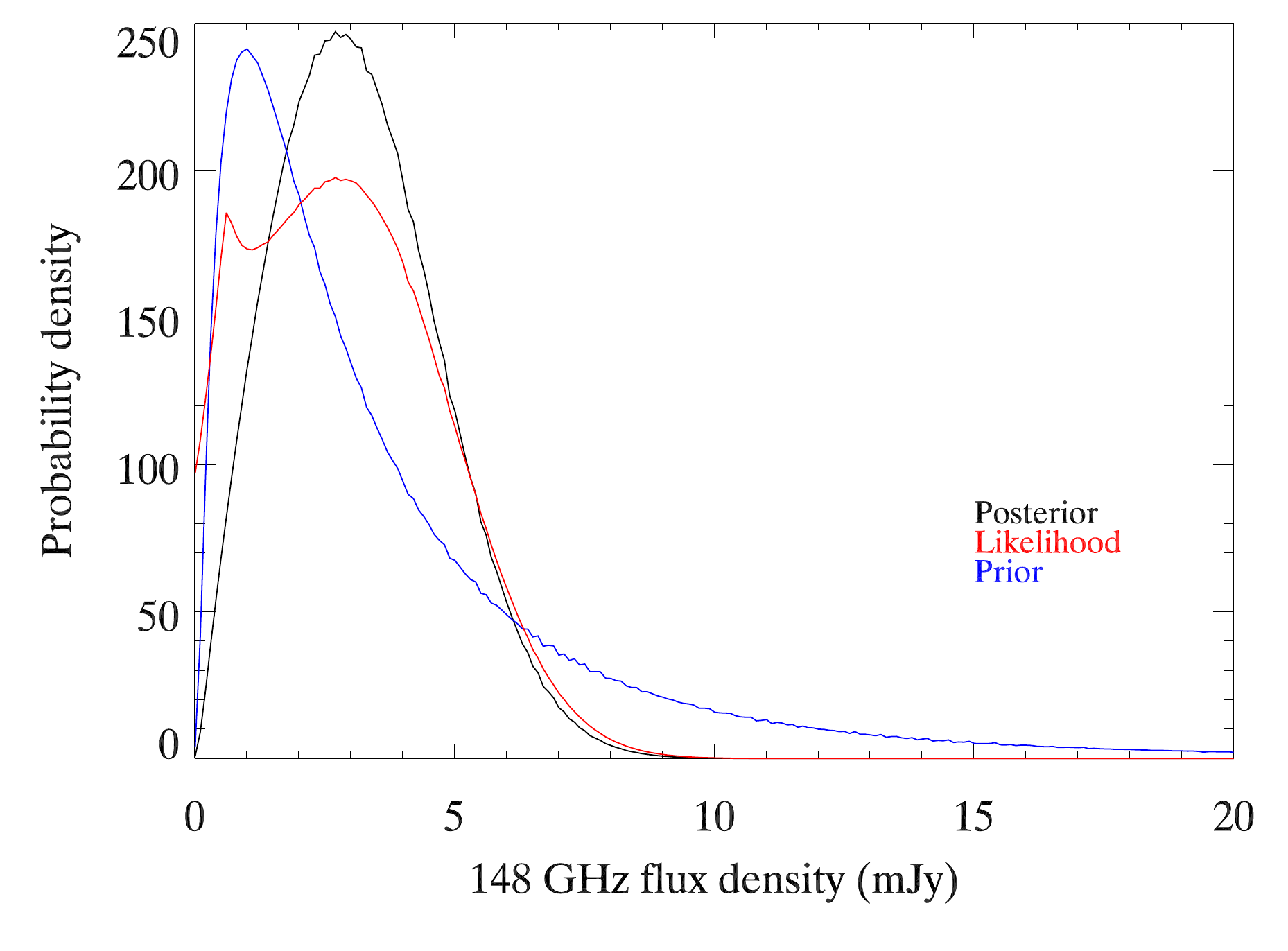}
\caption{The secondary-band posterior probability density function (black), the prior probability density function (blue), and the likelihood function (red), all expressed as a function of $S_{i,148}$. The posterior is calculated by multiplying the prior as a function of both $S_{i,218}$ and $\alpha$ by the likelihood similarly expressed, but all are shown here as functions of $S_{i, 148}$ for illustrative purposes only. \label{example1d}}
\end{figure}

To calculate the posterior distribution for the intrinsic 148~GHz flux density, we multiply each pixel of the prior (which is a function of $S_{i,218}$ and $\alpha$) by each pixel of the likelihood function (which similarly is calculated as a function of $S_{i,218}$ and $\alpha$). Figure \ref{example1d} shows the posterior distribution that results.  For each of these pixels, we calculate the 148~GHz flux density from the $S_{i,218}$ and $\alpha$.  We sum the probability density of all the $S_{i,218}$,$\alpha$ pixels that correspond to a given 148~GHz flux density (Equation \ref{eqn:bayesdebias2dto1d}). This procedure produces the posterior probability density function for the 148~GHz flux density, shown in Figure \ref{example1d}.  The median of the posterior distribution is reported as the debiased flux density at 148~GHz. For this example source, the debiased 148~GHz flux density is 3.1~mJy. The $68\%$ confidence interval of the posterior distribution for this example source ranges from 1.7 to 4.8~mJy.

\begin{figure*}[ht]
	\centering
\hfill
\subfigure[218 GHz, primary band]{\includegraphics[width=8.4cm]{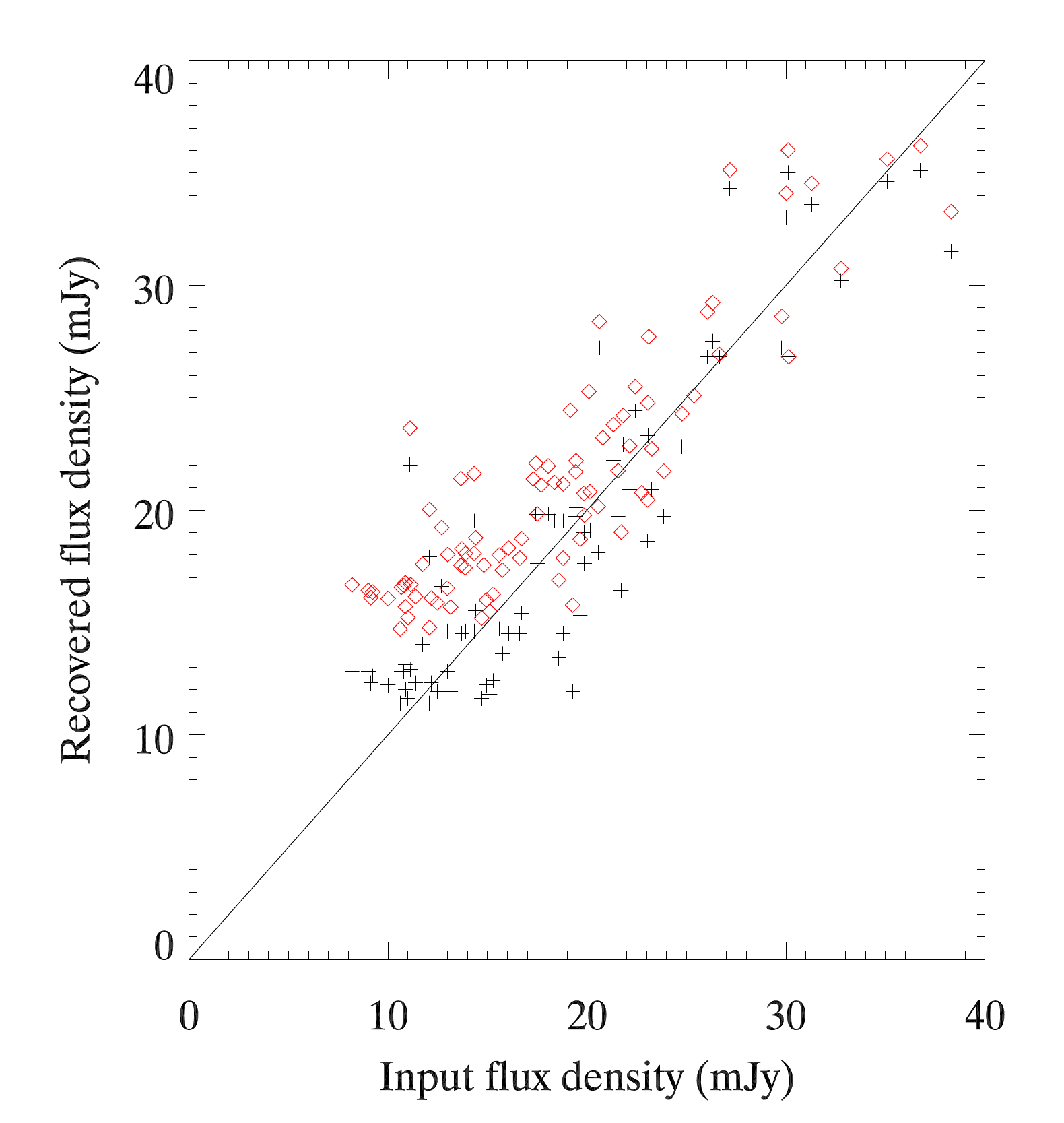}}
\hfill
\subfigure[148 GHz, secondary band]{\includegraphics[width=8.4cm]{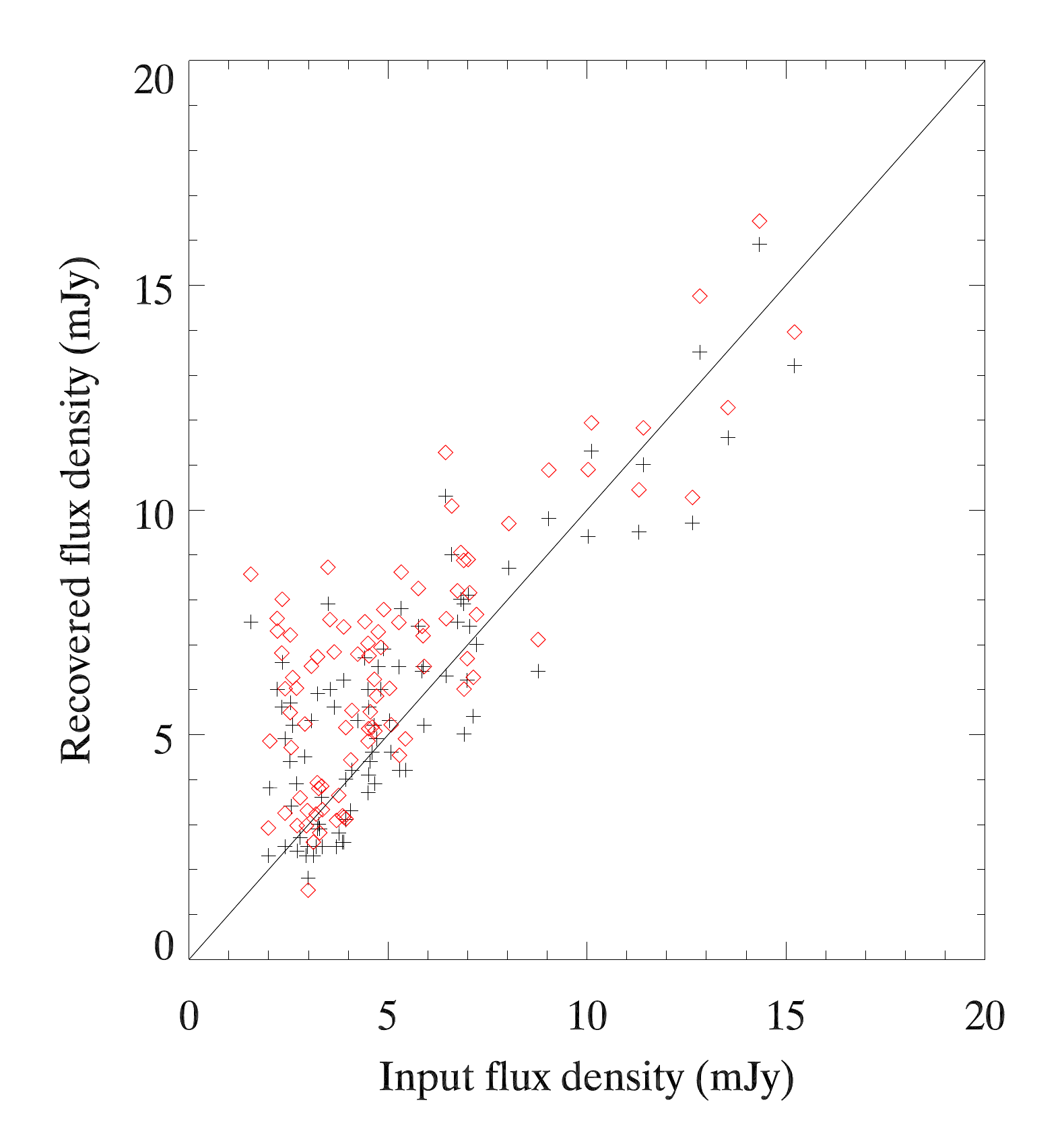}}
\hfill
\caption{The recovered flux densities as a function of input flux density for the simulated sample of dusty galaxies. The black crosses show the debiased flux densities as the recovered flux densities, and the red diamonds show the raw measured flux densities as the recovered flux densities.  The left plot shows the 218~GHz flux densities, which is the band at which these sources are selected. The right plot shows the 148~GHz flux densities, many of which are below the detection threshold of the ACT survey data. \label{fig:verifydebias}}
\end{figure*}
\subsection{Details}

The implementation of the multi-band debiasing methods requires a number of choices, which are described here in more detail.  These choices include selecting an adequate number of simulated sources to derive the likelihood function, selecting a radius for matching the recovered simulated sources with input sources, selecting the tolerance in flux density for selecting simulated sources with measured flux density close to that of the source being debiased, and selecting the kernel size of the kernel density estimator used to calculate the likelihood function.  

To test whether the simulated source sample is sufficiently large, we ran the debiasing on an example source (with 148 and 218 flux densities of 25~mJy) using only half of the simulated sources. The debiased flux densities agreed with those calculated from the entire simulated source sample, indicating that the number of sources in our simulations is sufficient for sampling the likelihood function.

To compute the flux density likelihood $P(S_m|S_i)$, we associate sources detected from the maps with input simulated sources. In so doing we must use an association radius. 
As discussed in \citet{grallasources}, the ACT positional uncertainty of sources with $S/N$ above 5 is typically less than 0.005~degrees. We adopt 0.007~degrees, or $25 \arcsec$, as our match radius. 
One of the simulated sources was randomly positioned close (within the 0.002~degree tolerance) to a real, bright source in the catalog, and it was excluded from the sample. We exclude cases in which two sources are measured to match an input source position. 

Our debiasing method requires that we identify sources from the simulations with flux densities similar to the catalog source that we are debiasing.  The default tolerance we use for 218~GHz selected sources is $\pm 1.0$~mJy. For 148~GHz selected sources and sources selected through our DSFG-optimized multi-frequency matched filter (MMF), we use a tolerance of $\pm 1.0$~mJy. For more details on 148~GHz and MMF selection, see the catalog paper \citep{grallasources}. 

At the low flux density end, the completeness affects the source recovery, such that the number of sources drops off very quickly toward low flux densities.  Because of the steepness of this completeness cut off, we must use a narrower tolerance when debiasing sources to avoid biasing the debiased flux density high, as lower flux density sources have not been detected.  Thus for sources with flux densities in the region of steeply falling completeness (below 25~mJy for 148~GHz selected sources), we reduce the tolerance to $\pm 0.5$~mJy.  To test whether the choice of tolerance has an significant effect, if we adopt $\pm0.5$~mJy as the tolerance, a value substantially less than the typical noise level ($\pm2.4$~mJy), the debiased flux densities do not change significantly compared to the $\pm1.0$~mJy tolerance. We note that this result is not surprising: the $\pm1.0$~mJy tolerance designates the limits of a uniform distribution, the effective standard error of which is $\pm 0.58$~mJy. Added in quadrature with the error on the likelihood, this choice only broadens the recovered posterior by 2\%.

The kernel density estimator we use to calculate the likelihood from the distribution of simulated sources requires specifying a scale for the kernel size, which can affect the results. Kernels that are too large would artificially smooth the likelihood function, inflating the tails of the distribution. Kernels that are too small would introduce noise into the likelihood function. After investigating a range of kernel scales, we set the kernel scale to 1.0~mJy. 

In addition to the choices discussed above, care must be taken when using the simulated source catalogs, particularly regarding spurious matches. For example, consider faint simulated sources that are spuriously matched with brighter real (or simulated) sources. When calculating the likelihood function of a source with measured flux densities like the brighter source, these spurious sources would contribute at low intrinsic flux densities. When this likelihood function is multiplied by the steep prior, this effect is amplified. Such sources must be vetoed from the simulations as they are not, by construction, the brightest source associated with a detection, which is what the likelihood is built to describe. For additional details of the application of these methods to the ACT data, such as the selection of the match radius and the handling of spurious matches, see \citet{grallasources}.

\subsection{Method applied to a population}

Finally, we used a second set of simulations to verify that the debiasing methods do indeed bring the flux densities into agreement with intrinsic flux densities. For this test, we populated the maps with sources with flux densities drawn according to the source counts model for the dusty source population, populating a range between 7 and 40~mJy.  The spectral index of each source was drawn from a Gaussian distribution with mean 3.4 and width 1.0, which is motivated by the measured spectral index distribution for dusty galaxies in the ACT catalog.  After performing the full source finding and flux density debiasing on this sample, we compare the input and debiased flux densities, shown in Figure \ref{fig:verifydebias}. As seen in this figure, the debiased flux densities improve upon the measured flux densities in recovering the input flux densities for this simulated dusty source sample. The median of the ratio of the input flux density to the debiased flux density is 0.98 for 218~GHz and 0.95 for 148~GHz. For comparison, the median of the ratio of the input flux density to the measured flux density is 0.88 for 218~GHz and 0.82 for 148~GHz.

\section{Conclusions}\label{sec:conclusion}

We have presented a new Bayesian method for debiasing source flux density estimates. The method combines an analytic prior with simulations for computing the likelihood of the intrinsic flux density. The simulations capture details of nontrivial source selection while the analytic prior provides a fast and flexible way to correct for flux bias arising from steeply falling counts. In the companion catalog paper \citep{grallasources}, we explore this method applied to DSFGs selected with a multifrequency matched filter with the additional complication of population culling to reduce contamination from Galactic sources. In this paper we have derived the method, provided a symbolic guide to implementation, and illustrated the steps applied to an example source. We have also shown how the method debiases the flux densities of a population of DSFGs with steeply falling number counts.

\section*{Acknowledgements}
We thank Eric Switzer, Tom Crawford, and Bruce Partridge for useful discussions and comments. 
This work was supported by the U.S. National Science Foundation through awards AST-0408698 and AST-0965625 for the ACT project, as well as awards PHY-0855887 and PHY-1214379. Funding was also provided by Princeton University, the University of Pennsylvania, and a Canada Foundation for Innovation (CFI) award to UBC. ACT operates in the Parque Astron\'omico Atacama in northern Chile under the auspices of the Comisi\'on Nacional de Investigaci\'on Cient\'ifica y Tecnol\'ogica de Chile (CONICYT). Computations were performed on the GPC supercomputer at the SciNet HPC Consortium. SciNet is funded by the CFI under the auspices of Compute Canada, the Government of Ontario, the Ontario Research Fund -- Research Excellence; and the University of Toronto. M.G. and T.M. acknowledge support from Johns Hopkins University. 

\bibliography{actequsources}
\bibliographystyle{apj}

\end{document}